\documentclass[12pt]{article}
\usepackage{epsf,latexsym}
\usepackage{amsfonts,amssymb} 
\epsfverbosetrue
\usepackage[ps,arc,frame]{xy}

\newcommand{\rxyn}[2]{{\begin{xy} 0;<2mm,0mm>:<0mm,2mm>::0;0,
,(5,-2)*{a}
,(10,-1.8)*{b}
,(15,-2)*{c}
,(20,-2)*{d}
,(25,-2)*{e}
,(30,-2)*{f}
,(2,-5)*{a}
,(2,-10)*{b}
,(2,-15)*{c}
,(2,-20)*{d}
,(2,-25)*{e}
,(2,-30)*{f}
,(5,-5)*\cir(#1,0){}
,(10,-5)*\cir(#1,0){}
,(15,-5)*\cir(#1,0){}
,(20,-5)*\cir(#1,0){}
,(25,-5)*\cir(#1,0){}
,(30,-5)*\cir(#1,0){}
,(5,-10)*\cir(#1,0){}
,(10,-10)*\cir(#1,0){}
,(15,-10)*\cir(#1,0){}
,(20,-10)*\cir(#1,0){}
,(25,-10)*\cir(#1,0){}
,(30,-10)*\cir(#1,0){}
,(5,-15)*\cir(#1,0){}
,(10,-15)*\cir(#1,0){}
,(15,-15)*\cir(#1,0){}
,(20,-15)*\cir(#1,0){}
,(25,-15)*\cir(#1,0){}
,(30,-15)*\cir(#1,0){}
,(5,-20)*\cir(#1,0){}
,(10,-20)*\cir(#1,0){}
,(15,-20)*\cir(#1,0){}
,(20,-20)*\cir(#1,0){}
,(25,-20)*\cir(#1,0){}
,(30,-20)*\cir(#1,0){}
,(5,-25)*\cir(#1,0){}
,(10,-25)*\cir(#1,0){}
,(15,-25)*\cir(#1,0){}
,(20,-25)*\cir(#1,0){}
,(25,-25)*\cir(#1,0){}
,(30,-25)*\cir(#1,0){}
,(5,-30)*\cir(#1,0){}
,(10,-30)*\cir(#1,0){}
,(15,-30)*\cir(#1,0){}
,(20,-30)*\cir(#1,0){}
,(25,-30)*\cir(#1,0){}
,(30,-30)*\cir(#1,0){}
#2\end{xy}}}

\newcommand{\double}[1]{\mathbb{#1}}

\newcommand{\cc}{\double{C}}

\newcommand{\rr}{\double{R}}
\newcommand{\zz}{\double{Z}}
\newcommand{\qqq}{\double{Q}}

\newcommand{\aaa}{\mathcal{A}}

\newcommand{\hhh}{\double{H}}
\newcommand{\mm}{\mathcal{M}}

\newcommand{\pp}{\pmatrix}

\newcommand{\ot}{\otimes}
\newcommand{\op}{\oplus}

\newcommand{\bb}{\begin{eqnarray}}
\newcommand{\ee}{\end{eqnarray}}
\newcommand{\eee}{\nonumber\end{eqnarray}}

\begin{document}

\font\twelve=cmbx10 at 13pt
\font\eightrm=cmr8

\thispagestyle{empty}

\begin{center}

Institut f\"ur Mathematik  $^3$ \\ Universit\"at Potsdam
\\ Am Neuen Palais 10 \\14469 Potsdam \\ Germany\\

\vspace{2cm}

{\Large\textbf{On a Classification of Irreducible \\
 Almost-Commutative
Geometries V}} \\

\vspace{1.5cm}

{\large Jan--Hendrik Jureit$^1$ \& Christoph A. Stephan$^{2,3}$}

\vspace{2cm}

{\large\textbf{Abstract}}
\end{center}
We extend a classification of irreducible, almost-commutative
geometries whose spectral action is dynamically non-degenerate, to
internal algebras that have six simple summands.

We find essentially four particle models: An extension of
the standard model by a new species of fermions with
vectorlike coupling to the gauge group and gauge invariant masses, two
versions of the electro-strong model and a variety 
of the electro-strong model with Higgs mechanism.

\vspace{2cm}

\noindent
PACS-92: 11.15 Gauge field theories\\
MSC-91: 81T13 Yang-Mills and other gauge theories

\vskip 1truecm

\noindent \\

\vspace{1.5cm}
\noindent $^1$ Privatgelehrter, Kiel (Germany) \\
\noindent $^2$ christophstephan@gmx.de\\

\newpage

\section{Introduction}

The classification presented here is based on the
formulation of the standard model in the language
of spectral triples \cite{con}, especially its recent
development \cite{barrett,cc} for spectral
triples in KO-dimension six.
We continue the classification of finite, real
spectral triples begun in \cite{1,2,3,Spinlift,4}.
So far spectral triples with up to four summands
in the finite matrix algebra have been considered, in
KO-dimension zero \cite{1,2,3,Spinlift} as well as
KO-dimension six \cite{4}.

Here we will investigate the case with six summands
in the matrix algebra of a finite spectral triple in KO-dimension
six. This classification is again with respect to a ``shopping
list'' consisting of heteroclitic criteria. For the exact
definitions we refer to \cite{1,4}.

Two criteria are motivated
from perturbative quantum field theory of the non-gravitational
part in flat timespace: vanishing Yang-Mills anomalies and
dynamical non-degeneracy. The later imposes that the number
of possible fermion mass equalities be restricted to the minimum and
that they be stable under renormalization flow. 

Two criteria are motivated from particle phenomenology: We want
the fermion representation to be complex under the little group in
each of its irreducible components, because we want to
distinguish particles from anti-particles by means of unbroken
charges. We want possible massless fermions to remain neutral under
the little group, i.e. not to couple to massless gauge bosons.

Two criteria are motivated from the hope that, one day, we will have
a unified quantum theory of all forces: vanishing mixed gravitational
Yang-Mills anomalies and again dynamical non-degeneracy.

Again heavy use of Krajewski diagrams was made \cite{kraj}. We
consider only minimal Krajewski diagrams. These were obtained
with help of an computer algebra program \cite{js} especially
developed for this task. This program already proved to be 
extremely useful in the previous classifications.

Unfortunately the computational time diverges rapidly when
increasing the number of summands in the matrix algebra. 
While for up to four summands only $\sim 20$ minimal
Krajewski diagrams appear the case of six summands
with 72 diagrams is much more complicated.
Luckily the diagrams fall into equivalent classes, i.e.
all possible relative direction of the arrows appear, and
so they allow a relatively compact notation by deleting
the arrow heads \cite{4}. But nevertheless all of the diagrams
have to be found and analysed; this becomes an
increasingly difficult terrain. 

Surprisingly the number of particles models remains 
rather small. We find again the electro-strong model
\cite{3} and two extensions of it.  But the prominent
role is played by a model which turns out to contain
the standard model as a sub-model. Considering
the number of minimal Krajewski diagrams this is
quite an astonishing result.

It would of course be desirable to understand the origin
of the internal space, i.e. the source of the matrix algebra.
There are hints that a connection to loop quantum gravity
exists \cite{dan}. Also  double Fell bundles
seem a plausible structure in noncommutative geometry \cite{ra1}. 
They could provide a deep connection to category theory
and give better insights into the mathematical structure of
almost-commutative geometries such as the standard model.

Another open problem is the mass mechanism for neutrinos.
In KO-dimension zero the masses are of Dirac type \cite{gracia,neutrino,ra2},
while KO-dimension six also allows for Majorana masses \cite{barrett,cc}
and the SeeSaw mechanism,
although  minor problems  concerning an axiom of
noncommutative geometry may occur \cite{ko6}. Another possibility 
lies in the modification of the spectral action \cite{sit}.

For a numerical analysis of the standard model with SeeSaw mechanism we refer
to \cite{cc,sm1,sm2} for the models with three and four summands
in the matrix algebra. 

\section{Statement of the result}

We consider a finite, real,  irreducible
spectral triple in KO-dimension six 
whose algebra has six simple summands
and the extended lift as described in \cite{fare}. Consider the list of all
Yang-Mills-Higgs models induced by these triples and lifts. Discard
all models that have either
\begin{itemize}\item
 a dynamically
degenerate fermionic mass spectrum,
\item
 Yang-Mills or gravitational
anomalies,
\item
 a fermion multiplet whose representation under
the little group is real or pseudo-real,
\item
or a massless fermion transforming
non-trivially under the little group.
\end{itemize}
 The remaining models are the
following, $A$ and $E$ is the number of colours, the gauge group is on the
left-hand side of the arrow, the little group on the right-hand side:
\begin{description}

\item[The standard model type:]
$A\geq 2$, $E\geq 2$
\bb
U(1) \times SU(2) \times SU(A) \times U(1) \times SU(E) &
\rightarrow
&
U(1)  \times SU(A) \times U(1) \times SU(E)
\eee
This model contains for $A=3$ one generation of 
the standard model in its four-summand version
\cite{3} and a vectorlike multiplet of new particles with $SU(E)$ colour.
For details see diagram 2. 

We have the following sub-models 
\bb
U(1) \times SO(2) \times SU(A) \times U(1) \times SU(E) &
\rightarrow
&
U(1)  \times SU(A) \times U(1) \times SU(E),
\eee
\bb
U(1) \times SU(2) \times SO(A) \times U(1) \times SU(E) &
\rightarrow
&
U(1)  \times SO(A) \times U(1) \times SU(E)
\eee
and
\bb
U(1) \times SU(2) \times SU(A)  \times SO(E) &
\rightarrow
&
U(1)  \times SU(A)  \times SO(E).
\eee
The last sub-model looses one of the U(1) subgroups since
only a linear combination couples to the fermions. It follows that the 
second $U(1)$ becomes unphysical and is not present 
in the spectral action.

If $A$ or $E$ are even we may also have the corresponding
symplectic groups as subgroups.
The sub-models contain  again each 
two sub-models where one of the special unitary 
groups is replaced by a special orthogonal or, if the
dimension is even, symplectic group. Note that in
this case there is always only one $U(1)$ coming
from the unitary group which is left.

\item[Electro-strong models without Higgs mechanism:]

These models do not contain a Higgs scalar and all the
fermions couple vectorially to the gauge group.
There are two models with the following gauge groups
\bb
U(1)\times SU(E)&
\rightarrow
&
U(1)\times SU(E),
\eee
\bb
U(1)\times U(1)\times SU(A)\times SU(E)&
\rightarrow
&
U(1)\times U(1)\times SU(A)\times SU(E)
\eee
and for the second case the obvious sub-models with orthogonal or symplectic
subgroups.

\item[Electro-strong models with Higgs mechanism:]

Here the $U(1)$ subgroup is broken through the Higgs mechanism:
\bb
U(1)\times SU(E)&
\rightarrow
&
\zz_2 \times SU(E)
\eee
In this model two of the fermion species have chiral $U(1)$ charges
while being $SU(E)$ singlets.
The fluctuations of the Dirac operator produce a
complex Higgs scalar. The third fermion species has vectorlike
couplings to the gauge group and its mass matrix is
gauge invariant. One finds that the minimum of the 
corresponding Higgs potential is non-degenerate.

\end{description}

\section{Diagram by diagram}

We will use the following letters to denote algebra elements and
unitaries: Let
$\aaa=M_A(\cc)\op M_B(\cc)\op M_C(\cc)\op M_D(\cc)
\op M_E(\cc) \op M_F(\cc) \owns
(a,b,c,d,e,f)$.
The extended lift is defined by
\bb 
L(u,v,w,x,y,z):=\rho (\hat u,\hat v,\hat w,\hat x,\hat y,\hat z)\,J \rho 
(\hat u,\hat v,\hat w,\hat x,\hat y,\hat z) J^{-1}
\ee
with
\bb 
\hat u&:=&u\,(\det u)^{q_{11}}(\det v)^{q_{12}}(\det
w)^{q_{13}}( \det x)^{q_{14}}( \det y)^{q_{15}}( \det z)^{q_{16}}\in U(A),\\
\hat v&:=&v\,(\det u)^{q_{21}}(\det v)^{q_{22}}(\det
w)^{q_{23}}( \det x)^{q_{24}}( \det y)^{q_{25}}( \det z)^{q_{26}}\in U(B),\\
\hat w&:=&w\,(\det u)^{q_{31}}(\det v)^{q_{32}}(\det
w)^{q_{33}}( \det x)^{q_{34}}( \det y)^{q_{35}}( \det z)^{q_{36}}\in U(C),\\
\hat x&:=&x\,(\det u)^{q_{41}}(\det v)^{q_{42}}(\det
w)^{q_{43}}( \det x)^{q_{44}}( \det y)^{q_{45}}( \det z)^{q_{46}}\in U(D),\\
\hat y&:=&y\,(\det u)^{q_{51}}(\det v)^{q_{52}}(\det
w)^{q_{53}}( \det x)^{q_{54}}( \det y)^{q_{55}}( \det z)^{q_{56}}\in U(E),\\
 \hat z&:=&z\,(\det u)^{q_{61}}(\det v)^{q_{62}}(\det
w)^{q_{63}}( \det x)^{q_{64}}( \det y)^{q_{65}}( \det z)^{q_{66}}\in U(F),,
\ee
with $q_{ij} \in \qqq$ 
and unitaries $(u,v,w,x,y,z)\in U( M_A(\cc)\op M_B(\cc)\op M_C(\cc)\op
M_D(\cc)\op M_E(\cc)\op M_F(\cc) )$. It is understood that for instance if $A=1$ we set
$u=1$ and $q_{j1}=0$,
$j=1,2,3,4,5,6$. If
$M_A(\cc)$ is replaced by $M_A(\rr)$ or $M_{A/2}(\hhh)$
 we set $q_{j1}=0$ and
$q_{1j}=0$.

To simplify the notation we introduce as in \cite{fare} the central charge matrix $Q$
\bb
Q:= ( q_{ij})
\ee 
where we do not spell out the  columns  which are identically zero due to 
the reasons just stated above. We will write the transpose $Q^t$
of the hypercharge matrix for obvious type setting reasons.

All the minimal Krajewski diagrams can be found at the end
of this paper.
The arrows in the Krajewski diagrams are not oriented to obtain a shorter 
notation. Each arrow is allowed to have any orientation with respect to the
other arrows. All minimal Krajewski diagrams for six summands in
the matrix algebra contain $2$ or $3$ arrows. 
It follows that each diagram encodes
$2$ or $2^2=4$ Krajewski diagrams with oriented arrows (the orientation
of one arrow can be fixed arbitrarily, only the relative orientations matter). 
\\ \\
{\bf Diagram 1}:
\\ \\
Since we are only dealing with minimal Krajewski diagrams we can arbitrarily 
choose an orientation for the arrows. All the other possible orientations are
implicitly covered by the choice of the representation of the algebra being
the fundamental or its complex conjugate. This would be no longer possible
if non-minimal Krajewski diagrams were considered since the orientation of the 
arrows, i.e. choosing which fermions are left-handed or right-handed, 
may allow for new mass terms in the fermionic mass matrix.  
For simplicity all arrows will  point into the same direction, namely to the
left.

Diagram 1 then takes its oriented form:

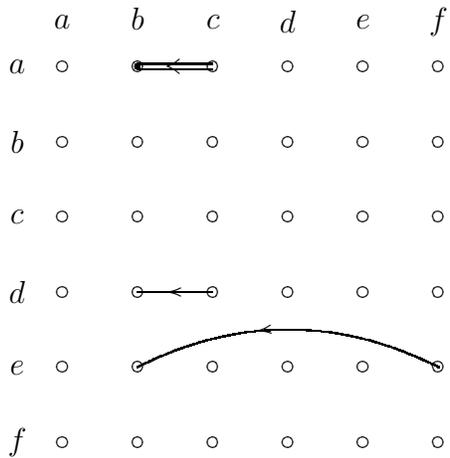
\begin{center}
\begin{tabular}{c}
\rxyn{0.4}{
,(10,-20);(15,-20)**\dir{-}?(.4)*\dir{<}
,(10,-5);(15,-5)**\dir2{-}?(.4)*\dir2{<}
,(10,-5)*\cir(0.2,0){}*\frm{*}
,(10,-25);(30,-25)**\crv{(20,-20)}?(.4)*\dir{<}
}
\end{tabular}
\end{center}

We read off the following representation for the algebra

\bb 
\rho _L =\pp{b\ot 1_A&0&0 \cr  0& ^{\beta_1} b\ot 1_D & 0 \cr 0 & 0&^{\beta_2} 
b\ot 1_E },&&
\rho _R=\pp{c\ot 1_A&0&0&0 \cr  0&^{\gamma_1} c \ot 1_A&0&0
\cr 0&0& ^{\gamma_2} c \ot 1_D &0 \cr 0&0&0& f \ot 1_E },\cr \cr \cr
\rho _L^c=\pp{1_B \ot  a&0 &0 \cr  0&1_B \ot d & 0 \cr 0&0& 1_B \ot e},&&
\rho _R^c=\pp{1_C\ot a &0&0&0 \cr  0&1_C \ot a &0&0
\cr 0&0& 1_C \ot d &0 \cr 0&0&0& 1_F \ot e },
\ee
 and the mass matrix:
\bb 
\mm=\pp{M_1\ot 1_A& M_2\ot 1_A & 0 &0\cr 0&  0&M_3 \ot1_D &0
\cr 0&  0&0& M_4 \ot1_D  },\cr \cr \cr  M_1, M_2, M_3 \in
M_{B\times
C}(\cc), \ M_4\in M_{B\times F }(\cc).
\ee
As in \cite{Spinlift} the 
parameters $\beta_1,\beta_2 $ and $\gamma_1,\gamma_2$ take values 
$\pm 1$ and
distinguish between fundamental representation and its complex
conjugate: $^1 b:=b$, $^{-1}b:=\bar b$.
The colour algebras consist of $a$s, $d$s and $e$s. To obtain a
non-degenerate mass spectrum we must impose $B=1$ otherwise
we would obtain at least two or more massless states (this procedure
was named ``neutrino counting'' in \cite{1} and proved to be an extremely
efficient tool to identify degenerate mass spectra). Additionally
these massless fermions would in general be charged under the little group,
so models with $B\geq 2$ fail at least in two ways to meet our
constraints. Furthermore, if $B=1$
anomaly cancellation imposes $A=1$, $C=1$ and $F=1$. 

So we are left with a model with two unbroken colour algebras $M_{D}(\cc)$
and $M_{E}(\cc)$. The unitaries of these two matrix algebras 
are to be lifted to the Hilbert space and the central extension of the
lift is required to be free of anomalies.

Let us assume that $D\geq 2$ and $E\geq 2$, so there are two $U(1)$'s
in the central extension.

For $(\beta_1,\beta_2,\gamma_1,\gamma_2)=(+,+,+,+),(+,-,+,+),(-,+,+,-),
(-,-,+,-)$ we find the following central charge matrix for the anomaly free central charges:
\bb
Q^t = \pp{q_{14}=-p &q_{24}=p &q_{34}=p &q_{44}=q  &q_{54}=r &q_{64}=p \cr 
q_{16}=-p &q_{26}=p &q_{36}=p &q_{46}=s &q_{56}=t &q_{66}=p } 
\ee
with $p,q,r,s,t \in \qqq $. All fermions possess vectorlike hypercharges,
so the  gauge group $U(1) \times U(1) \times SU(D) \times SU(E)$
is unbroken.

But the fermions represented by the double 
arrow in the $a$-line of the Krajewski diagram have vanishing hypercharge
and are neutral under the little group. 

All the other possibilities for $(\beta_1,\beta_2,\gamma_1,\gamma_2)$ 
contain also at least one irreducible chiral lepton with vanishing hypercharge.

The cases $D=1$, $E\geq 2$ and $D\geq 2$, $E=1$ have just one
colour algebra and consequently just one $U(1)$ subgroup in the 
gauge group. But as before the anomaly free lift implies at least one
lepton with vanishing hypercharge for all possible configurations
of $(\beta_1,\beta_2,\gamma_1,\gamma_2)$.

Note that the Krajewski diagram 1 may be enlarged to become
an anomaly free model with the standard model as a sub-model.
It is then no longer a minimal Krajewski diagram but it provides
an interesting model beyond the standard model, closely related
to Okun's theta particle model \cite{Okun}.
This extension has been studied in detail in \cite{newcolour}
\\ \\
{\bf Diagram 2}:
\\ \\
We take again the Krajewski diagram with all arrows pointing to the left:

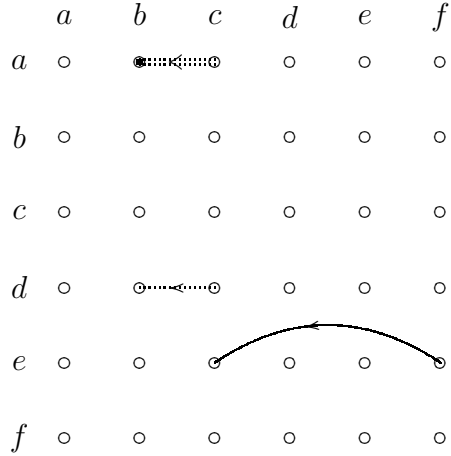
\begin{center}
\begin{tabular}{c}
\rxyn{0.4}{
,(10,-20);(15,-20)**\dir{..}?(.4)*\dir{<}
,(10,-5);(15,-5)**\dir2{..}?(.4)*\dir2{<}
,(10,-5)*\cir(0.2,0){}*\frm{*}
,(15,-25);(30,-25)**\crv{(22.5,-20)}?(.4)*\dir{<}
}
\end{tabular}
\end{center}

Here the dotted arrows indicate the sub-diagram which can 
produce the standard model as a sub-model.

We read off the following representation for the algebra:

\bb 
\rho _L =\pp{b\ot 1_A&0&0 \cr  0& ^{\beta} b\ot 1_D & 0 \cr 0 & 0& c \ot 1_E },&&
\rho _R=\pp{^{\gamma_1}  c\ot 1_A&0&0&0 \cr  0&^{\gamma_2} c \ot 1_A&0&0
\cr 0&0& ^{\gamma_3} c \ot 1_D &0 \cr 0&0&0& f \ot 1_E },\cr \cr \cr
\rho _L^c=\pp{1_B \ot  a&0 &0 \cr  0&1_B \ot d & 0 \cr 0&0& 1_C \ot e},&&
\rho _R^c=\pp{1_C\ot a &0&0&0 \cr  0&1_C \ot a &0&0
\cr 0&0& 1_C \ot d &0 \cr 0&0&0& 1_F \ot e },
\ee
 and for the mass matrix:
\bb 
\mm=\pp{M_1\ot 1_A& M_2\ot 1_A & 0 &0\cr 0&  0&M_3 \ot1_D &0
\cr 0&  0&0& M_4 \ot1_E  },\cr \cr \cr  M_1, M_2, M_3 \in
M_{B\times
C}(\cc), \ M_4\in M_{C\times F }(\cc).
\ee
Anomaly cancellation, a non-degenerate mass spectrum and
unbroken colour enforce $C=1$ and $F=1$.
Furthermore the constraint of a  non-degenerate mass spectrum leaves 
us with two possibilities for $B$, namely $B=1$ and $B=2$. 

The first possibility, $B=1$, produces the same models as diagram 1.

The second possibility, $B=2$, enforces $D=1$ to evade massless particles
charged under the little group. So there are two colour algebras, 
$M_A(\cc)$ which is the usual colour algebra of the quarks and
$M_E(\cc)$ which is the colour algebra of the new vectorlike leptons
corresponding to $M_4$, we call them X-particles.  
Altogether we have therefore three $U(1)$ 
subgroups in the gauge group. 

Let us first consider the configurations $(\beta,\gamma_1,\gamma_2,\gamma_3)
=(+,-,+,+), (+,+,-,+),$ $(-,-,+,+), (-,+,-,+), (+,-,+,-), (+,+,-,-), (-,-,+,-), (-,+,-,-)$. 

If $A\geq 2$ and $E \geq 2$ we find the following two central charge matrices
which provide for anomaly free lifts:
\bb
Q^t_1 = \pp{q_{11}=p &q_{21}=0 &q_{31}=p\, A -1 &q_{41}=-(p\, A -1) 
&q_{51}=p &q_{61}=p\, A -1 \cr 
q_{12}=q &q_{22}=\frac{1}{2} &q_{32}=q \, A &q_{42}= - q \, A &q_{52}= r 
&q_{62}= q \, A \cr 
q_{15}=p - \frac{1}{A}  &q_{25}=0  &q_{35}=p \, A -1 &q_{45}= -(p \, A -1) &
q_{55}=s  &q_{65}=p \, A -1 } 
\label{qsm1}
\ee
\bb
Q^t_1 = \pp{q_{11}=p &q_{21}=0 &q_{31}=p\, A -1 &q_{41}=-(p\, A -1) 
&q_{51}=p &q_{61}=p\, A -1 \cr 
q_{12}=q &q_{22}=\frac{1}{2} &q_{32}=q \, A &q_{42}= - q \, A &q_{52}= r 
&q_{62}= q \, A \cr 
q_{15}=0  &q_{25}=0  &q_{35}=0 &q_{45}=0 &
q_{55}=s  &q_{65}=0 }
\label{qsm2}
\ee
with $p,q,r,s \in \qqq $. For the standard model sub-model the 
hypercharges of all three $U(1)$s are proportional. Therefore
we can choose our basis, as in the case of the pure standard
model \cite{3}, in such a way that only one linear combination
of the three $U(1)$s  is relevant to the standard model particles.
We call this linear combination $U(1)_Y$ since it constitutes
the hypercharge group. The new X-particles corresponding
to $M_4$ are charged under all three $U(1)$s. Rotating first
to the standard model basis leaves us with one more rotation
in the X-particle sector, so effectively the new particles
are charged under $U(1)_Y$ and a second $U(1)_X$. One
linear combination of the three $U(1)$s always decouples 
and is not present in the spectral action. 

The standard model sector exhibits the same Higgs mechanism
as the pure standard model \cite{cc} while the X-particles 
couple vectorially to the gauge group and have therefore
gauge invariant masses. So the gauge group is broken
and we have:
\bb
U(1)_Y \times SU(2) \times SU(A) \times U(1)_X \times SU(E)
\rightarrow U(1)_{em} \times SU(A) \times U(1)_X \times SU(E)
\ee
Note that the sub-representation on the standard model
Hilbert space still has the non-trivial kernel $\zz_2\times \zz_A$.
But these elements of $U(1)_Y$ are not necessarily in the
kernel of the sub-representation on the X-particle
Hilbert space.

Finally, if $A=1$ or $E=1$ or both we find anomaly free
lifts which correspond to simply setting the central
charges $q_{j1}$ or $q_{j5}$ in the charge matrices
(\ref{qsm1}) and (\ref{qsm2}) to zero.

For $A\geq2$, $E=1$ we find 
\bb
U(1)_Y \times SU(2) \times SU(A) \times U(1)_X 
\rightarrow U(1)_{em} \times SU(A) \times U(1)_X, 
\ee
This model is very similar to the AC-model presented
in \cite{beyond} which contains two new particles of
the X-type.
For $A=1$, $E\geq 2$
\bb
U(1)_Y \times SU(2)  \times U(1)_X \times SU(E)
\rightarrow U(1)_{em}  \times U(1)_X \times SU(E)
\ee
which is an extension of the electro-weak model 
in \cite{3}.
If $A=1$ and $E=1$ we have
\bb
U(1)_Y \times SU(2)
\rightarrow U(1)_{em}, 
\ee
because in this case only one $U(1)$ remains to be lifted.

The eight other possibilities  $(\beta,\gamma_1,\gamma_2,\gamma_3)$
lead for any $A$ and $E$ to  models
containing at least one irreducible chiral lepton with vanishing hypercharge.
\\ \\
{\bf Diagrams 3, 4 and 6} produce the same models as diagram 2. 
The arrows of the standard model particles can be easily identified
using the list of Krajewski diagrams in \cite{sms}.
\\ \\
{\bf Diagram 5} behaves as diagram 1.
\\ \\
{\bf Diagram 7} yields with our standard orientation for the arrows:
\\
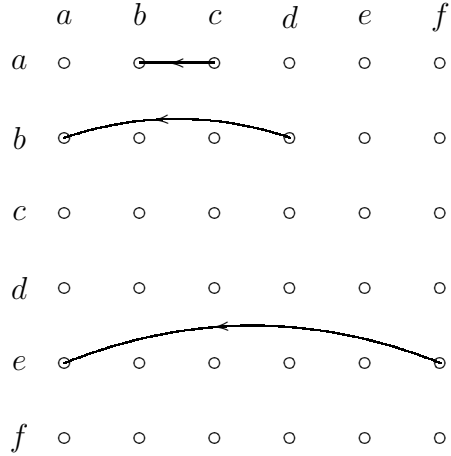
\begin{center}
\begin{tabular}{c}
\rxyn{0.4}{
,(10,-5);(15,-5)**\dir{-}?(.4)*\dir{<}
,(5,-10);(20,-10)**\crv{(12.5,-7.5)}?(.4)*\dir{<}
,(5,-25);(30,-25)**\crv{(17.5,-20)}?(.4)*\dir{<}
} 
\end{tabular}
\end{center}
The representation of the algebra is then found to be
\bb 
\rho _L =\pp{b\ot 1_A&0&0 \cr  0&  a \ot 1_B & 0 \cr 0 & 0& ^{\alpha} a \ot 1_E },&&
\rho _R=\pp{ c\ot 1_A&0&0 \cr  0&d \ot 1_B&0 \cr 0&0& f \ot 1_E },\cr \cr \cr
\rho _L^c=\pp{1_B \ot  a&0 &0 \cr  0&1_A \ot \, ^{\beta}b & 0 \cr 0&0& 1_A \ot e},&&
\rho _R^c=\pp{1_C\ot a &0&0 \cr  0&1_D \ot \, ^{\beta}b &0
\cr 0&0& 1_F \ot e  },
\ee
 and the mass matrix we find
\bb 
\mm=\pp{M_1\ot 1_A& 0 &0\cr 0&  M_2 \ot1_B &0
\cr  0&0& M_3 \ot1_E  },\cr \cr \cr  M_1\in M_{B\times C}(\cc), \ M_2 \in M_{A\times D }(\cc), \ M_3 \in M_{A\times F }(\cc) .
\ee
To obtain models with unbroken colour algebras that are free of 
massless particles charged under the little group, we have
to choose  $A=B=C=D=F=1$. Furthermore $E\geq 2$ and  $M_E(\cc)$ 
has to be over the complex numbers since otherwise there would
be no $U(1)$ to be lifted.

For $(\alpha_1,\alpha_2,\beta)=(+,+,+),(-,+,+)$ we find the
anomaly free charge matrices
\bb
Q^t_1= \pp{q_{15}=-q & q_{25}=q & q_{35}=-p & q_{45}=p & q_{55}=r & q_{65}=-q } 
\ee
and
\bb
Q^t_2= \pp{q_{15}=p & q_{25}=q & q_{35}=q & q_{45}=p & q_{55}=r & q_{65}=p }, 
\ee
with $p,q,r \in \qqq$.
While $Q_1$ has irreducible leptons with vanishing hypercharge, $Q_2$ 
provides an extension of the electro-strong model \cite{3} by one
set of particles coupling vectorially to the gauge group. The  
unbroken gauge group is
\bb
U(1) \times SU(E) \, \rightarrow U(1) \times SU(E).
\ee
For $(\alpha_1,\alpha_2,\beta)=(+,-,+),(+,+,-),(-,-,+),(-,+,-)$ we
find three anomaly free charge matrices:
\bb
Q^t_1= \pp{q_{15}=p & q_{25}=q & q_{35}=q & q_{45}=p & q_{55}=r & q_{65}=p },
\ee
\bb
Q^t_2= \pp{q_{15}=-q & q_{25}=p & q_{35}=-2q+p & q_{45}=p & q_{55}=r & q_{65}=-p },
\ee
\bb
Q^t_3= \pp{q_{15}=p & q_{25}=0 & q_{35}=p-q & q_{45}=q & q_{55}=r & q_{65}=p }
\ee
with $p,q,r \in \qqq$. $Q_1$ and $Q_2$ produce the same extension of the
electro-strong model as above.

The charge matrix $Q_3$ leads to chiral couplings and consequently
we get, by virtue of the fluctuations of the mass matrices $M_1$ and
$M_2$, a complex Higgs scalar $h$,
\bb
\varphi_1 = \sum_i r_i (\det y)^{-(p+q)} M_1= M_1 \, h, \quad \varphi_2 = \sum_i r_i (\det y)^{p+q} M_2 = M_2 \, \bar{h},  \quad r_i \in \rr ,
\ee
for which the Higgs potential has a non-degenerate minimum. The third mass
$M_3$ is gauge invariant, similarly  to the electro-strong model.
The Higgs mechanism breaks the gauge group
\bb
U(1) \times SU(E) \rightarrow \zz_2  \times SU(E).
\ee
Here we have included the finite subgroup $\zz_2$ which we 
usually do not spell out explicitly.
The remaining possibilities for $(\alpha_1,\alpha_2,\beta)$
have the anomaly free charge matrix
\bb
Q^t= \pp{q_{15}=-p+q+r & q_{25}=p & q_{35}=q & q_{45}=r & q_{55}=s & 
q_{65}=-p+q+r }
\ee
with $p,q,r,s \in \qqq$. This leads again to chiral U(1) couplings and
we find the same Higgs mechanism as above with  
\bb
U(1) \times SU(E) \rightarrow \zz_2  \times SU(E),
\ee
where again the fluctuations of $M_1$ and $M_2$ produce the
complex Higgs scalar, while $M_3$ is a gauge invariant mass term.
\\ \\
{\bf Diagrams 8,9 and 10} produce the same models as diagram 7.
\\ \\
{\bf Diagram 11} yields:

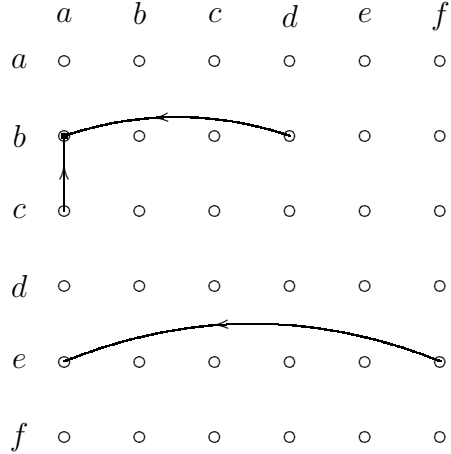
\begin{center}
\begin{tabular}{c}
\rxyn{0.4}{
,(5,-10)*\cir(0.2,0){}*\frm{*}
,(5,-10);(5,-15)**\dir{-}?(.4)*\dir{<}
,(5,-10);(20,-10)**\crv{(12.5,-7.5)}?(.4)*\dir{<}
,(5,-25);(30,-25)**\crv{(17.5,-20)}?(.4)*\dir{<}
}
\end{tabular}
\end{center}
From the Krajewski diagram we read off the representation
\bb 
\rho _L =\pp{a\ot 1_B&0  \cr  0&  ^{\alpha_1} a \ot 1_E  },&&
\rho _R=\pp{^{\alpha_2} a \ot 1_C &0&0 \cr  0&d \ot 1_B&0 \cr 0&0& f \ot 1_E },\cr \cr \cr
\rho _L^c=\pp{1_A \ot  b&0 \cr  0&1_A \ot  e },&&
\rho _R^c=\pp{1_A \ot c &0&0 \cr  0&1_D \ot b &0
\cr 0&0& 1_F \ot e  },
\ee
 and the mass matrix
\bb 
\mm=\pp{ 1_A \ot M_1 &  M_2 \ot 1_B  &0\cr 0&  0 & M_3 \ot 1_E},
\cr \cr \cr  M_1\in M_{B\times C}(\cc), \ M_2 \in M_{A\times D }(\cc), \ M_3 \in M_{A\times F }(\cc) .
\ee
The models produce either massless particles charged under the little
group, anomalous models or possess particles with vanishing charge.
\\ \\
{\bf Diagram 12} shares the fate of diagram 11.
\\ \\
{\bf Diagram 13} yields:

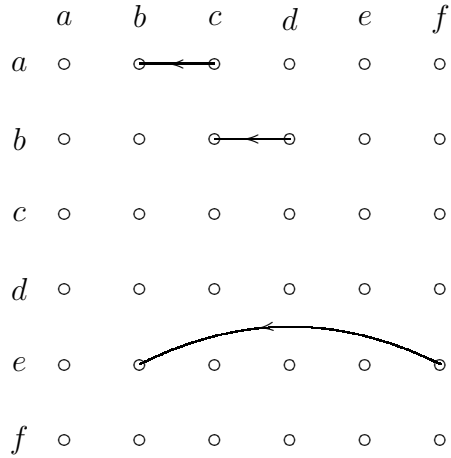
\begin{center}
\begin{tabular}{c}
\rxyn{0.4}{
,(10,-5);(15,-5)**\dir{-}?(.4)*\dir{<}
,(15,-10);(20,-10)**\dir{-}?(.4)*\dir{<}
,(10,-25);(30,-25)**\crv{(20,-20)}?(.4)*\dir{<}
}
\end{tabular}
\end{center}
We find the representation
\bb 
\rho _L =\pp{b\ot 1_A&0&0 \cr  0&  c \ot 1_B & 0 \cr 0 & 0& b \ot 1_E },&&
\rho _R=\pp{ c \ot 1_A&0&0 \cr  0&d \ot 1_B&0 \cr 0&0& f \ot 1_E },\cr \cr \cr
\rho _L^c=\pp{1_B \ot \,  a&0 &0 \cr  0&1_C \ot \, ^{\beta} b & 0 \cr 0&0& 1_B \ot \, e},&&
\rho _R^c\pp{1_C \ot \,  a&0 &0 \cr  0&1_D \ot \, ^{\beta} b & 0 \cr 0&0& 1_F \ot \, e},
\ee
and the mass matrix
\bb 
\mm=\pp{M_1\ot 1_A& 0 &0\cr 0&  M_2 \ot1_B &0
\cr  0&0& M_3 \ot1_E  },\cr \cr \cr  M_1\in M_{B\times C}(\cc), \ M_2 \in M_{C\times D }(\cc), \ M_3 \in M_{B\times F }(\cc) .
\ee
A non-degenerate mass spectrum, unbroken colours and the absence  
of harmful anomalies requires $B=C=D=F=1$ and $A \geq 2$ or $E \geq 2$
being complex matrix algebras. If either $A=1$ or $E=1$ the
diagram is treated as diagram 7.

If both $A \geq 2$ and $E\geq 2$ and complex then we find an anomaly
free lift with charge matrix
\bb
Q^t= \pp{q_{11}=p & q_{21}=q & q_{31}=q & q_{41}=q & q_{51}=r & 
q_{61}=q  \cr
q_{15}=s & q_{25}=t & q_{35}=t & q_{45}=t & q_{55}=l & 
q_{65}=t },
\ee
$p,q,r,s,t,l \in \qqq$,
where the hypercharges and colour groups couple vectorially. This result
holds for all $(\beta)$.
Consequently all three mass matrices are gauge invariant and the gauge
group is unbroken:
\bb
U(1) \times U(1) \times SU(A) \times SU(E) \rightarrow 
U(1) \times U(1) \times SU(A) \times SU(E).
\ee 
We have therefore an electro-strong model with two colour groups and
two hypercharge groups. Note that the representation does not
allow to choose a linear combination that decouples one $U(1)$ subgroup
as in the standard model.
\\ \\
{\bf Diagrams 14, 15 and 16} produce the same models as diagram 13.
\\ \\
{\bf Diagram 17} has no unbroken colour and thus no $U(1)$ to lift.
\\ \\
{\bf Diagram 18}:
This diagram is treated the same way as diagram 2 with the
restriction $A=1$. 
\\ \\
{\bf Diagram 19} gives the same models as diagram 1, i.e. at least
on fermion has vanishing charge.

\section{Conclusion and outlook}

For three and four summands in the matrix algebra we essentially
found only the standard model and the electro-strong model
as compatible models for our ``shopping list'', \cite{1,2,3,4}.  
Of course there
were a few sub-models due to the fact that the gauge group
has orthogonal or, if the dimension permits, symplectic
subgroups. These sub-models exist  also in the
extension of the standard model found in this paper.

Here we considered the case of six summands, which again
provided a relatively sparse list of models. Apart from
the electro-strong model, possibly enlarged by an
extra colour group and with the possibility of a Higgs mechanism,
the standard model took again its prominent place. 
It comes as a sub-model and is accompanied by an extra species 
of fermions, the X-particles 
which couple vectorially to the standard model gauge group.
The standard model gauge group appears as a subgroup
of a larger gauge group since the X-particles are allowed to possess
their own colour group.
This new model is a smaller version of the 
already known AC-model \cite{beyond} which has proven to
provide for an interesting dark matter candidate \cite{khlop}.

Taking into account that we have 72 minimal 
Krajewski diagrams for finite spectral triples
with six summands in the matrix algebra
(if all possible arrow configurations
are counted), it is quite astonishing to find so few
different particle models.

An other surprising result is the empirical fact that
all possible arrow orientations of a given minimal
Krajewski diagram also constitute a minimal Krajewski
diagram. This seems only to be true in $KO$-dimension
six since there exist counterexamples in $KO$-dimension
zero \cite{1,2,3}. A general proof for any number
of summands in the internal algebra would be desirable.
It would greatly reduce the difficulties of finding
all minimal Krajewski diagrams for a given number
of summands in the internal algebra.

Again we are curious to know what happens with
eight (and more) summands. A viable approach seems
a restriction to extensions which contain the standard model
as sub-model. This may keep the computational time reasonable.

\subsection*{Acknowledgements}

The author gratefully acknowledges the funding of his work
by the Deutsche Forschungsgemeinschaft.

\vfil\eject
\enlargethispage{1cm}
\thispagestyle{empty}

\begin{figure}
\begin{center}
\begin{tabular}{ccc}
\rxyn{0.4}{
,(10,-20);(15,-20)**\dir{-}
,(10,-5);(15,-5)**\dir2{-}
,(10,-5)*\cir(0.2,0){}*\frm{*}
,(10,-25);(30,-25)**\crv{(20,-20)}
}
&
\quad \quad \quad \quad
&
\rxyn{0.4}{
,(10,-20);(15,-20)**\dir{-}
,(10,-5);(15,-5)**\dir2{-}
,(10,-5)*\cir(0.2,0){}*\frm{*}
,(15,-25);(30,-25)**\crv{(22.5,-20)}
} 
\\ \\
Diagram 1 && Diagram 2
\\ \\
\rxyn{0.4}{
,(10,-20);(15,-20)**\dir{-}
,(10,-5);(15,-5)**\dir2{-}
,(10,-5)*\cir(0.2,0){}*\frm{*}
,(20,-25);(30,-25)**\crv{(25,-22)}
}
&
\quad \quad \quad \quad
&
\rxyn{0.4}{
,(10,-5)*\cir(0.2,0){}*\frm{*}
,(10,-5);(15,-5)**\dir2{-}
,(10,-20);(25,-20)**\crv{(17.5,-17)}
,(15,-30);(25,-30)**\crv{(20,-27.5)}
}
\\ \\
Diagram 3 && Diagram 4
\\ \\
\rxyn{0.4}{
,(10,-5)*\cir(0.2,0){}*\frm{*}
,(10,-5);(15,-5)**\dir{-}
,(10,-5);(20,-5)**\crv{(15,-7.5)}
,(10,-25);(15,-25)**\dir{-}
,(10,-30);(20,-30)**\crv{(15,-27.5)}
}
&
\quad \quad \quad \quad
&
\rxyn{0.4}{
,(10,-5)*\cir(0.2,0){}*\frm{*}
,(10,-5);(15,-5)**\dir{-}
,(10,-5);(20,-5)**\crv{(15,-7.5)}
,(10,-25);(15,-25)**\dir{-}
,(15,-30);(20,-30)**\dir{-}
}
\\ \\
Diagram 5 && Diagram 6
\end{tabular}
\end{center}
\end{figure}

\begin{figure}
\begin{center}
\begin{tabular}{ccc}
\rxyn{0.4}{
,(10,-5);(15,-5)**\dir{-}
,(5,-10);(20,-10)**\crv{(12.5,-7.5)}
,(5,-25);(30,-25)**\crv{(17.5,-20)}
}
&
\quad \quad \quad \quad
&
\rxyn{0.4}{
,(10,-5);(15,-5)**\dir{-}
,(5,-10);(20,-10)**\crv{(12.5,-7.5)}
,(15,-25);(30,-25)**\crv{(22.5,-22)}
}
\\ \\
Diagram 7 && Diagram 8
\\ \\
\rxyn{0.4}{
,(10,-5);(15,-5)**\dir{-}
,(15,-10);(20,-10)**\dir{-}
,(5,-25);(30,-25)**\crv{(17.5,-20)}
}
&
\quad \quad \quad \quad
&
\rxyn{0.4}{
,(10,-5);(15,-5)**\dir{-}
,(20,-10);(25,-10)**\dir{-}
,(15,-20);(30,-20)**\crv{(22.5,-17)}
}
\\ \\
Diagram 9 && Diagram 10
\\ \\
\rxyn{0.4}{
,(5,-10)*\cir(0.2,0){}*\frm{*}
,(5,-10);(5,-15)**\dir{-}
,(5,-10);(20,-10)**\crv{(12.5,-7.5)}
,(5,-25);(30,-25)**\crv{(17.5,-20)}
}
&
\quad \quad \quad \quad
&
\rxyn{0.4}{
,(5,-10)*\cir(0.2,0){}*\frm{*}
,(5,-10);(5,-15)**\dir{-}
,(5,-10);(20,-10)**\crv{(12.5,-7.5)}
,(15,-25);(30,-25)**\crv{(22.5,-22)}
}
\\ \\
Diagram 11 && Diagram 12
\end{tabular}
\end{center}
\end{figure}

\begin{figure}
\begin{center}
\begin{tabular}{ccc}
\rxyn{0.4}{
,(10,-5);(15,-5)**\dir{-}
,(15,-10);(20,-10)**\dir{-}
,(10,-25);(30,-25)**\crv{(20,-20)}
}
&
\quad \quad \quad \quad
&
\rxyn{0.4}{
,(10,-5);(15,-5)**\dir{-}
,(20,-10);(25,-10)**\dir{-}
,(15,-30);(20,-30)**\dir{-}
}
\\ \\
Diagram 13 && Diagram 14
\\ \\
\rxyn{0.4}{
,(10,-5);(15,-5)**\dir{-}
,(15,-10);(20,-10)**\dir{-}
,(15,-25);(30,-25)**\crv{(22.5,-22)}
}
&
\quad \quad \quad \quad
&
\rxyn{0.4}{
,(10,-5);(15,-5)**\dir{-}
,(15,-10);(20,-10)**\dir{-}
,(20,-25);(30,-25)**\crv{(25,-22.5)}
}
\\ \\
Diagram 15 && Diagram 16
\\ \\
\rxyn{0.4}{
,(10,-5);(15,-5)**\dir{-}
,(20,-10);(25,-10)**\dir{-}
,(5,-20);(30,-20)**\crv{(17.5,-15)}
}
&
\quad \quad \quad \quad
&
\rxyn{0.4}{
,(10,-20);(15,-20)**\dir{-}
,(10,-5);(15,-5)**\dir2{-}
,(10,-5)*\cir(0.2,0){}*\frm{*}
,(5,-30);(25,-30)**\crv{(15,-25)}
}
\\ \\
Diagram 17 && Diagram 18
\end{tabular}
\end{center}
\end{figure}

\begin{figure}
\begin{center}
\begin{tabular}{c}
\rxyn{0.4}{
,(5,-10)*\cir(0.2,0){}*\frm{*}
,(5,-10);(5,-15)**\dir{-}
,(5,-10);(20,-10)**\crv{(12.5,-7.5)}
,(10,-25);(15,-25)**\dir{-}
,(5,-30);(20,-30)**\crv{(12.5,-25)}
} 
\\ \\
Diagram 19
\end{tabular}
\end{center}
\end{figure}

\enlargethispage{1cm}


\begin{thebibliography}{30}

\bibitem{con}
 A. Connes, {\it Noncommutative Geometry}, Academic Press (1994)\\
 A. Connes, {\it Noncommutative geometry and
reality},  J. Math. Phys. 36 (1995) 6194\\
A. Connes \& M. Marcolli {\it Noncommutative Geometry, Quantum Fields and Motives }
(2007) http://www.alainconnes.org/ \\
A. Connes, {\it Gravity coupled with matter and the
foundation of noncommutative geometry}, hep-th/9603053, Comm.
Math. Phys. 155 (1996) 109\\
 A. Chamseddine \& A. Connes, {\it The spectral action principle},
hep-th/9606001, Comm. Math. Phys. 182 (1996) 155
\bibitem{barrett}
J. Barrett,
{\it A Lorentzian version of the non-commutative geometry of the  standard model of particle physics}, hep-th/0608221,
J. Math. Phys. 48 (2007)012303
\bibitem{cc}
A. Connes, {\it Noncommutative geometry and the standard model with  neutrino mixing}, hep-th/0608226, JHEP 0611 (2006) 081
\\
A. Chamseddine, A. Connes \& M. Marcolli,
{\it Gravity and the standard model with neutrino mixing},
hep-th/0610241
\bibitem{1}
B. Iochum,  T. Sch\"ucker \& C.A. Stephan,  {\it On a classification of
irreducible almost commutative geometries}, hep-th/0312276,
J. Math. Phys.  {45} (2004) 5003
\bibitem{2}
J.-H. Jureit \&  C.A. Stephan,  {\it On a classification of
irreducible almost commutative geometries, a second helping}, hep-th/ 0501134,
J.\ Math.\ Phys.\  {46} (2005) 043512
\bibitem{3}
J.-H. Jureit, T. Sch\"ucker \&  C.A. Stephan,  {\it On a classification of
irreducible almost commutative geometries III}, hep-th/0503190,
J. Math. Phys.  {46} (2005) 072303
\bibitem{Spinlift}
T. Sch\"ucker, {\it
Krajewski diagrams and spin lifts},
hep-th/0501181
\bibitem{4}
J.-H. Jureit \&  C.A. Stephan,  {\it On a classification of
irreducible almost commutative geometries IV}, hep-th/0610040,
J.Math.Phys. {49} (2008) 033502
\bibitem{kraj}
M. Paschke \& A. Sitarz, {\it Discrete spectral triples and
their symmetries}, q-alg/9612029, J. Math. Phys. 39 (1998) 6191\\
T. Krajewski, {\it Classification of finite spectral triples},
hep-th/9701081, J. Geom. Phys. 28 (1998) 1
\bibitem{js}
J.-H. Jureit \& C.A. Stephan,
{\it  Finding the standard model of particle physics: A Combinatorial problem},
hep-th/0503085,
Comput.Phys.Commun. {178} (2008) 230-247
\bibitem{dan}
 J. Aastrup \& J. Grimstrup, {\it Spectral triples of holonomy loops},
hep-th/0503246, Commun. Math.  Phys.  264 (2006) 657  \\
J.Aastrup \& J. Grimstrup,
{\it Intersecting Connes noncommutative geometry with quantum gravity},
hep-th/0601127, Int. J. Mod. Phys. A  22 (2007) 1589 \\
J.Aastrup, J.Grimstrup \& R.Nest,
{\it On Spectral Triples in Quantum Gravity I},
arXiv:0802.1783 [hep-th] \\
J.Aastrup, J. Grimstrup \& R.Nest,
{\it On Spectral Triples in Quantum Gravity II},
arXiv:0802.1784 [hep-th] \\
J.Aastrup, J.Grimstrup \& R.Nest,
{\it A new spectral triple over a space of connections},
arXiv:0807.3664 [hep-th]
\bibitem {ra1}
R.A.D. Martins,
{\it Double Fell bundles and Spectral triples},
arXiv:0709.2972 [math-ph] \\
R.A.D. Martins,
{\it Some constructions in Category theory and Noncommutative geometry},
arXiv:0811.1485 [math-ph]
\bibitem{gracia}
J.M. Gracia-Bondia,
{\it Connes' interpretation of the standard model and massive neutrinos},
arXiv:hep-th/9502120, Phys. Lett. B 351 (1995)
\bibitem{neutrino}
C.A. Stephan,
{\it Massive neutrinos in almost commutative geometry},
arXiv:hep-th/0608053,  J. Math. Phys. 438 (2007) 023513
\bibitem{ra2}
J. W. Barrett  \& R. A. Martins, 
{\it  Non-commutative geometry and the standard model vacuum},
hep-th/0601192, J.Math.Phys. {47} (2006) 052305 \\
R.A.D. Martins, 
{\it  Finite temperature corrections and embedded strings in noncommutative 
geometry and the standard model with neutrino mixing}
arXiv:0705.0613 [hep-th], J.Math.Phys. {48} (2007) 083509
\bibitem{ko6}
C.A. Stephan, {\it Almost-commutative geometry, massive neutrinos and the
orientability axiom in KO-dimension 6}, hep-th/0610097 (2006)
\bibitem{sit}
A. Sitarz, {\it Spectral action and neutrino mass}, arXiv:0808.4127 [math-ph] (2008)
\bibitem{sm1}
J.-H. Jureit, T. Krajewski, T. Sch\"ucker \& C.A. Stephan,
 {\it Seesaw and noncommutative geometry},
arXiv:0801.3731 [hep-th],  Phys. Lett. B 654 (2007) 127 
\bibitem{sm2}
J.-H. Jureit, T. Krajewski, T. Schucker \& C.A. Stephan,
{\it On the noncommutative standard model},
arXiv:0705.0489 [hep-th],  Acta Phys. Polon. B 38 (2007) 3181 
\bibitem{fare}
 S. Lazzarini \& T. Sch\"ucker, {\it A farewell to
unimodularity}, hep-th/0104038,  Phys.Lett. B 510 (2001) 277
\bibitem{Okun}
L.B. Okun, {\it Theta particles}, Nucl. Phys. B 173 (1980) 1
\bibitem{newcolour}
C.A. Stephan, 
{\it  Almost-commutative geometries beyond the standard model II. New Colours}
arXiv:0706.0595 [hep-th]
J. Phys. A 40 (2007) 9941
\bibitem{beyond}
C.A. Stephan,
{\it Almost-commutative geometries beyond the standard model}, hep-th/ 0509213
J. Phys. A39 (2006) 9657
\bibitem{sms}
C.A. Stephan,
{\it Krajewski diagrams and the Standard Model},
arXiv:0809.5137 [hep-th]
\bibitem{khlop}
D. Fargion, M. Khlopov \& C.A. Stephan,
{\it Cold dark matter by heavy double charged leptons?},
astro-ph/0511789, Class. Quant. Grav. 23 (2006) 7305\\
M. Y. Khlopov \& C.A. Stephan,
{\it Composite dark matter with invisible light from almost-commutative
geometry}, astro-ph/0603187
\end{thebibliography}
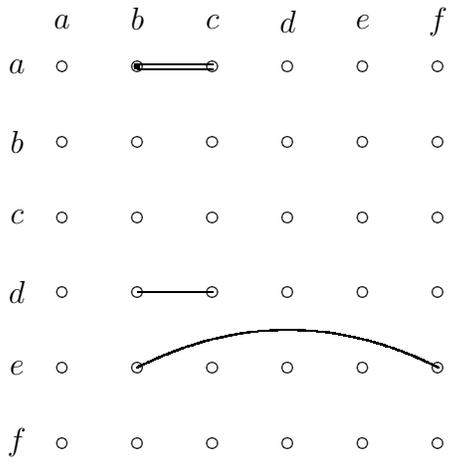
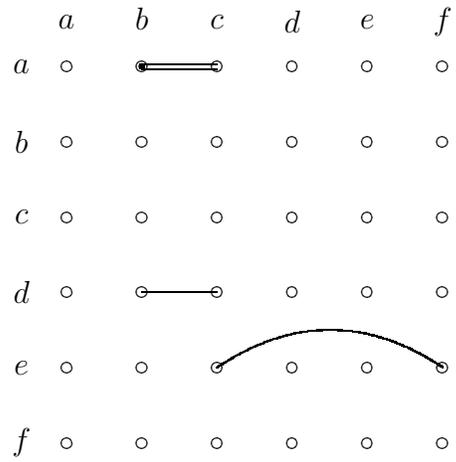
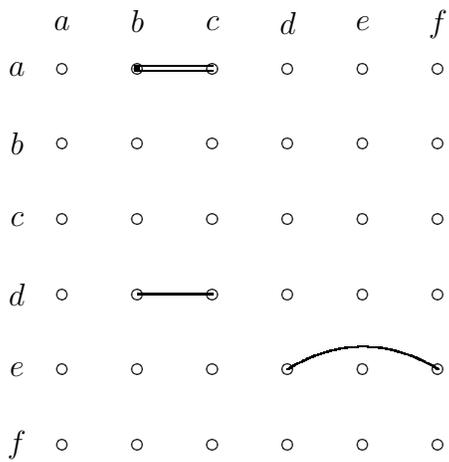
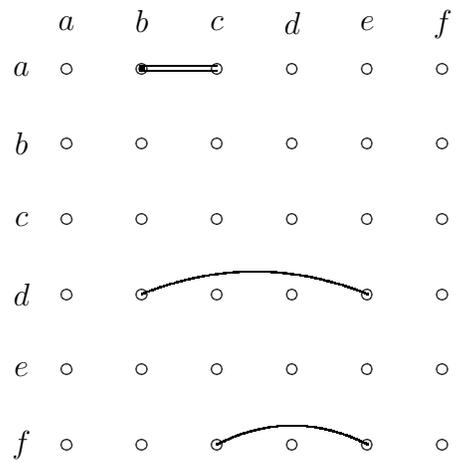
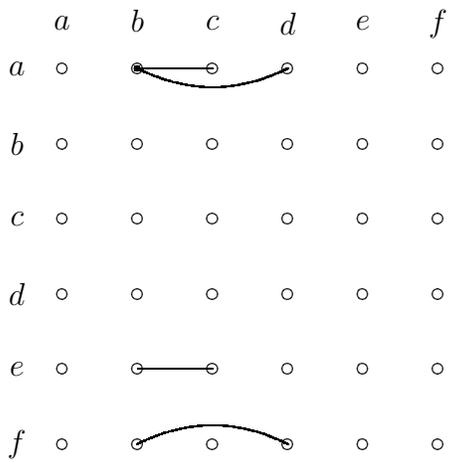
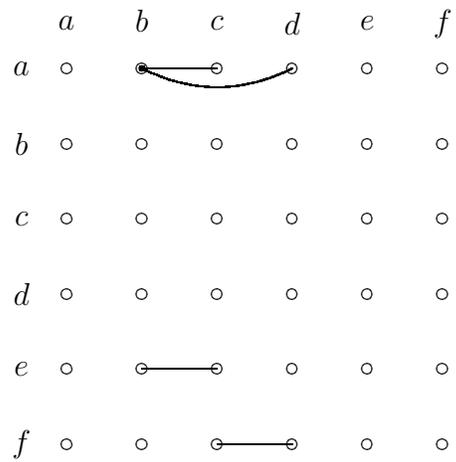
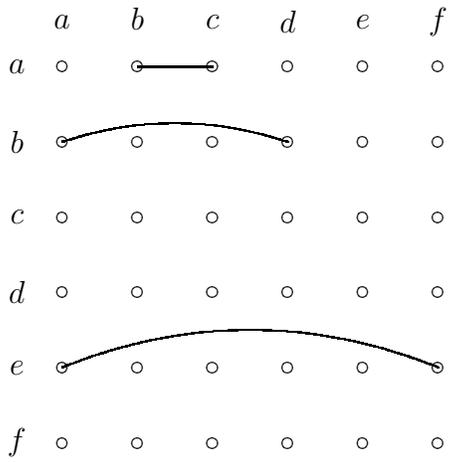
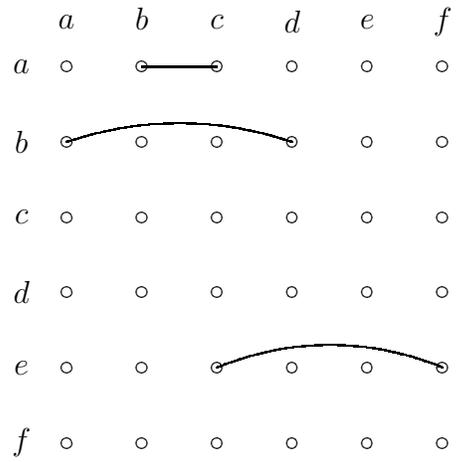
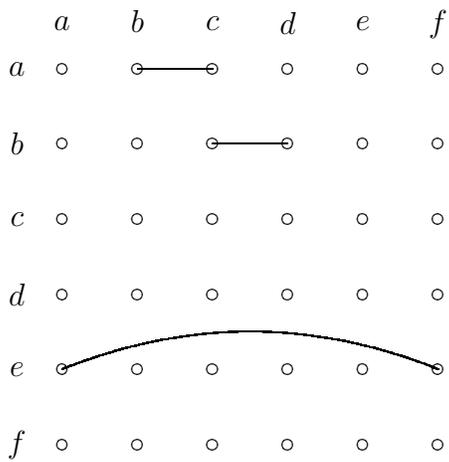
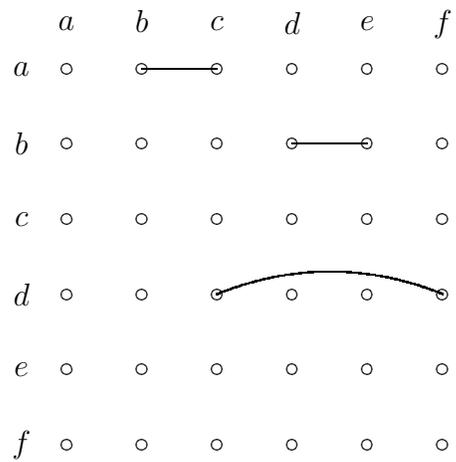
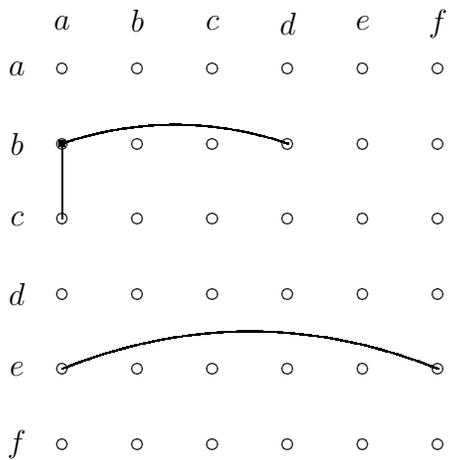
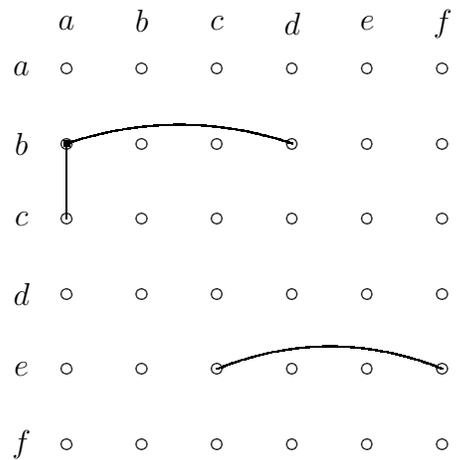
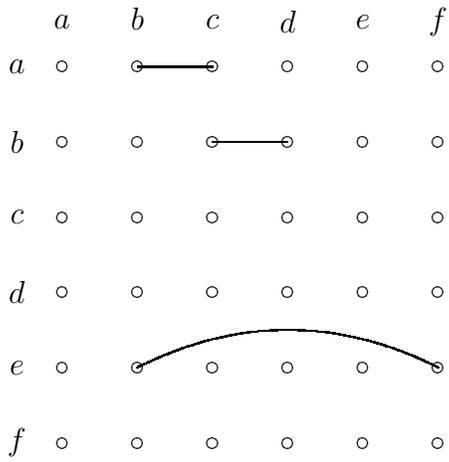
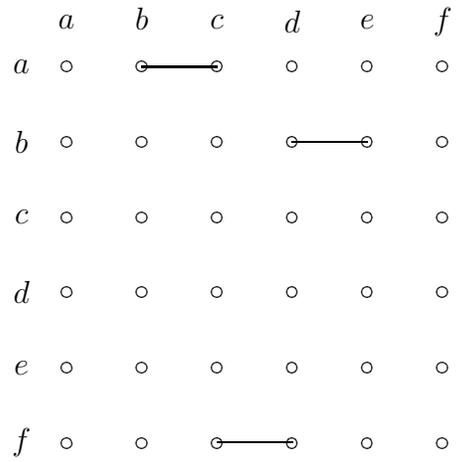
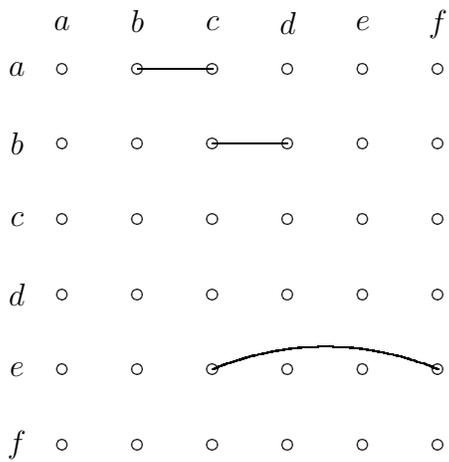
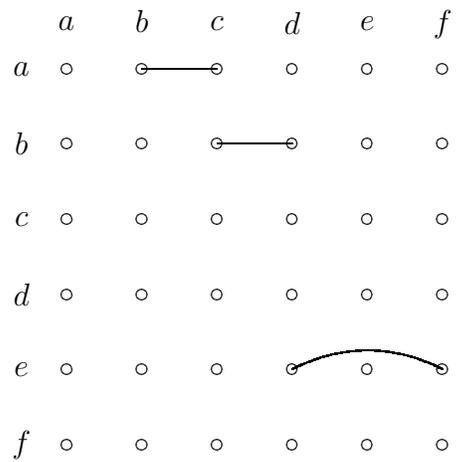
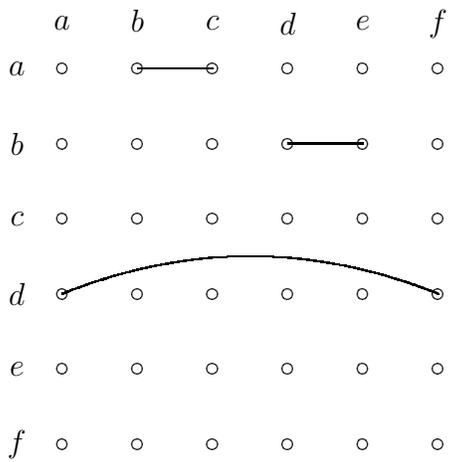
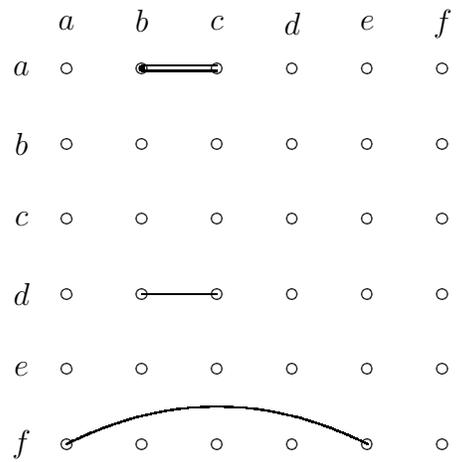
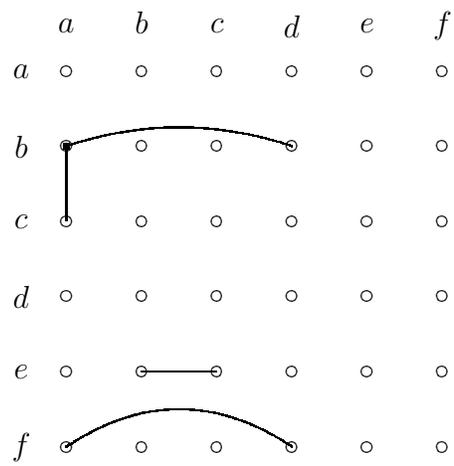
\end{document}